\documentclass[cameraready]{Interspeech}

\title{Towards Interpretable Framework for Neural Audio Codecs via Sparse Autoencoders: A Case Study on Accent Information}

\author[affiliation={1}]{Shih-Heng}{Wang}
\author[affiliation={1}]{Tiantian}{Feng}
\author[affiliation={1}]{Aditya}{Kommineni}
\author[affiliation={1}]{Thanathai}{Lertpetchpun}
\author[affiliation={1}]{Bowen}{Yi}
\author[affiliation={1}]{Xuan}{Shi}
\author[affiliation={1}]{Shrikanth}{Narayanan}

\address{
    $^1$ University of Southern California, USA 
}

\email{shihheng@usc.edu}


\usepackage{comment}
\usepackage{amsmath}
\usepackage{amssymb}
\usepackage{mathrsfs}
\usepackage{xcolor} 
\usepackage{booktabs}
\usepackage{makecell}
\usepackage{cite}

\begin{document}

\maketitle



\begin{abstract}
Neural Audio Codecs (NACs) are widely adopted in modern speech systems, yet how they encode linguistic and paralinguistic information remains unclear. Improving the interpretability of NAC representations is critical for understanding and deploying them in sensitive applications. Hence, we employ Sparse Autoencoders (SAEs) to decompose dense NAC representations into sparse, interpretable activations. In this work, we focus on a challenging paralinguistic attribute—accent—and propose a framework to quantify NAC task-level interpretability. We evaluate 4 four NAC models under 16 SAE configurations using a relative performance index. Our results show that DAC and SpeechTokenizer achieve the highest interpretability. We further reveal that acoustic-oriented NACs encode accent information primarily in activation magnitudes of sparse representations, whereas phonetic-oriented NACs rely more on activation positions, and that low-bitrate EnCodec variants show higher interpretability.
\end{abstract}

\vspace{-2mm}
\section{Introduction}
Neural Audio Codec (NAC) representations~\cite{soundstream,encodec,dac,speechtokenizer,hificodec,wu2024ts3,mousavi2025discrete,arora2025landscape,moshi,espnet_codec,survey,codecsuperb,dasb} have been widely adopted in modern speech processing systems, such as text-to-speech and speech foundation models, serving as compact discrete representations of speech~\cite{valle,valle2,vallex,moshi,audiolm,soundstorm}. 
This widespread usage has resulted in efforts towards benchmarking NAC representations on both speech understanding and generation tasks~\cite{codecsuperb,dasb,espnet_codec,survey}.
Additionally, preliminary works have investigated the robustness~\cite{codec_probe} and generalization ability~\cite{codec_generalize} of NAC representations.
While these works provide us with understanding of NACs' performance on varied speech tasks, the internal mechanisms of NACs—namely, how the learnt representations encode linguistic and paralinguistic information—remain underexplored.
Limited understanding of NAC internal mechanisms constrains the development of interpretable models for deployment in sensitive domains. In areas such as healthcare and assistive technologies, interpretability is critical for trustworthy decision-making.

To address this gap, we propose employing Sparse Autoencoders (SAEs)~\cite{sae1} to improve the interpretability of NAC representations.
Recent works in Natural Language Processing (NLP) have used SAEs to extract human-interpretable features from language models' representations~\cite{sae1,sae_claude,sae_gemma,topksae,superposition,sae_moral}. SAEs decompose polysemantic representations into sparse, high-dimensional activations where each dimension corresponds to an monosemantic feature.
In speech and audio domains, SAEs have been used to analyze ASR models~\cite{sae_asr,sae_attn_emotion,sae_mmi_asr,sae_phone}, foundation models for speech and music~\cite{sae_foundation,sae_audio,saemusic}, revealing task-relevant and acoustically meaningful features. 
Another recent study~\cite{sae_codec} shows that SAEs trained on latent representations of audio generative models can isolate interpretable acoustic properties such as pitch, amplitude and timbre.



As a first step towards interpreting NAC representations, we focus on \textbf{accent} information. Accent provides a challenging case for interpretability: it arises from interactions among speaker characteristics, phonetic realization, and contextual variation across utterances~\cite{acc_syth}. After compressed by NACs, these factors are further entangled within discrete codes, making interpreting how accent information is encoded difficult. By examining accent as an initial case study, we aim to establish a measurable framework for NAC interpretability, which can be extended to other speaker traits and paralinguistic attributes.

To achieve this goal, we develop a framework to measure NAC interpretability in accent information. Specifically, we focus on task-level interpretability in this work. 
We first extract NAC representations (Sec.~\ref{subsec:NAC}) and train SAEs to obtain interpretable sparse activations (Sec.~\ref{subsec:SAE}). 
Logistic regression (LR) classifiers are trained on these activations to perform accent classification (Sec.~\ref{subsec:LR}).
To further analyze how accent information is encoded, we propose to decompose the activations into two complementary components—\textbf{position} and \textbf{magnitude}—for analysis (Sec.~\ref{subsec:LR}). Last, for fair comparison, we propose to measure interpretability using the relative performance index $\Delta$F1 (Sec.~\ref{subsec:explain_measure}), defined with respect to each NAC reference performance.

\begin{figure*}[t]
    \centering
    \includegraphics[width=0.95\linewidth]{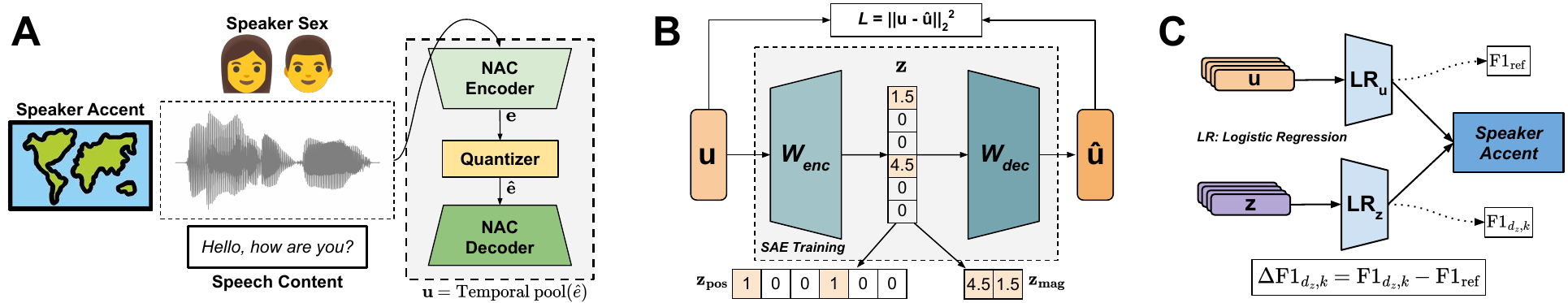}
    \vspace{-2mm}
    \caption{Sparse Autoencoder (SAE) framework for measuring interpretability of NACs. (A) Extraction of utterance level representation for a given speech input using NAC. (B) SAE training for learning sparse representation $\mathbf{z}$. Position ($\mathbf{z}_\text{pos}$) and Magnitude ($\mathbf{z}_\text{mag}$) are then derived from sparse representation. (C) Logistic regression models are trained for binary accent classification tasks, with the utterance level representation ($\mathbf{u}$) providing reference score $\text{F1}_{ref}$ and sparse representation corresponding to $\text{F1}_{d_z,k}$}
    \label{fig:sae}
    \vspace{-5mm}
\end{figure*}

In our experiments, we build two binary English accent classification tasks (Sec.~\ref{subsec:dataset}) with data described in the Vox-Profile benchmark~\cite{voxprofile}. 
We train SAE models with four NAC models—EnCodec~\cite{encodec}, DAC~\cite{dac}, Mimi~\cite{moshi}, and SpeechTokenizer~\cite{speechtokenizer}(Sec.~\ref{subsec:nac_setup}).
Each NAC is tested under 16 SAE configurations of varying latent size and sparsity (Sec.~\ref{subsec:sae_setup}). 
Our contributions and results are summarized as follows:
\begin{enumerate}
    \item We propose a framework for NAC task-level interpretability by learning SAE sparse representations and measuring their effectiveness through relative performance index, $\Delta$F1.
    \item DAC and SpeechTokenizer show the best accent-based interpretability among the four NACs.
    \item We observe different behaviors among NACs: acoustic-oriented NACs encode accent information primarily in activation magnitudes of sparse representations, whereas phonetic-oriented NACs rely more on activation positions.
\end{enumerate}

\vspace{-2mm}
\section{Framework}
\vspace{-1mm}
In this section, we will introduce our proposed framework and how we utilize it to measure the interpretability level of NACs. In Figure~\ref{fig:sae}, we give an overview of our framework.

\vspace{-2mm}
\subsection{Neural audio codec representations}
\label{subsec:NAC}
\vspace{-2mm}
Neural Audio Codecs (NACs)~\cite{soundstream, encodec, dac, speechtokenizer, hificodec, wu2024ts3, mousavi2025discrete, arora2025landscape, moshi} typically consist of three components during inference: an encoder, vector quantizers~\cite{vqvae}, and a decoder. Given a waveform, the encoder converts it into the continuous representation $\mathbf{e}$. Next, the vector quantizers quantize $\mathbf{e}$ into discrete codes $\mathbf{c}$, which can be used as speech representations for language modeling and other downstream tasks~\cite{valle,valle2,vallex}. Last, discrete codes $\mathbf{c}$ are mapped back to a continuous representation $\hat{\mathbf{e}} \in \mathbb{R}^{T \times d_\text{codec}}$, where $T$ represents time, and the decoder reconstructs the waveform with $\hat{\mathbf{e}}$. 

In Figure~\ref{fig:sae}A, we illustrate how NAC representations are incorporated into our framework. We apply mean pooling over time on $\hat{\mathbf{e}}$ to obtain an utterance-level representation $\mathbf{u} \in \mathbb{R}^{d_\text{codec}}$, as accent information distributes across the entire utterance. We analyze and train the SAE on the representation $\mathbf{u}$.

\vspace{-2mm}
\subsection{Sparse autoencoder}
\label{subsec:SAE}
\vspace{-2mm}
A Sparse Autoencoder (SAE) is used to decompose accent information from the utterance-level representations $\mathbf{u}$ as shown in Figure~\ref{fig:sae}B. While there are several SAE variants~\cite{sae1, topksae, gatesae}, in this study we adopt TopK SAE, used in prior speech SAE literature~\cite{sae_foundation}. 
This variant controls sparsity by retaining only the $k$ largest non-zero activations. 
Specifically, the SAE model consists of a linear encoder and a linear decoder with weights $W_\text{enc}$ and $W_\text{dec}$, respectively, without bias terms in either layer.
During training, $\mathbf{u}$ is first shifted by a bias term $b_\text{pre}$ and then projected into a higher-dimensional latent space. A top-$k$ activation is applied to enforce sparsity in the latent representation:
\begin{align}
    \mathbf{z} = \text{TopK}(W_\text{enc}(\mathbf{u} - b_\text{pre} ), k), \quad \mathbf{z} \in \mathbb{R}^{d_\text{z}}
    \label{eq:sae_e}
\end{align}
where $d_\text{z} > d_\text{codec}$. Next, the decoder reconstructs the input representation from the sparse activation $\mathbf{z}$, i.e., $\hat{\mathbf{u}} = W_\text{dec}\mathbf{z} + b_\text{pre}$. Finally, we update the model with the reconstruction loss using mean squared error (MSE), defined as $L = \lVert \mathbf{u} - \hat{\mathbf{u}} \rVert_{2}^{2}$. Additional implementation details can be found in~\cite{topksae}.

Within our framework, we use the learned sparse representation $\mathbf{z}$ for accent classification. This design is motivated by two considerations. First, $\mathbf{z}$ preserves information to reconstruct $\mathbf{u}$, ensuring that task-relevant information is retained. Second, $\mathbf{z}$ is sparser and more interpretable than the original representation $\mathbf{u}$, allowing us to probe not only whether accent information is present, but how it is organized within the representation.


\vspace{-2mm}
\subsection{Information encoding and logistic regression}
\label{subsec:LR}
\vspace{-2mm}
To analyze how $\mathbf{z}$ encodes accent information, as shown in Figure~\ref{fig:sae}B, we decompose $\mathbf{z}$ into two complementary components: \textbf{position} and \textbf{magnitude} features, denoted as $\mathbf{z}_\text{pos}$ and $\mathbf{z}_\text{mag}$.




\noindent \textbf{Position feature} $\mathbf{z}_\text{pos} \in \mathbb{R}^{d_\text{z}}$ preserves the indices of activated dimensions. Specifically, in $\mathbf{z}_\text{pos}$, only activated positions are assigned a value of 1, and all other positions are set to 0.

\noindent \textbf{Magnitude feature} $\mathbf{z}_\text{mag} \in \mathbb{R}^{k}$, in contrast, retains only the activation magnitudes of the top-$k$ active dimensions, sorted in descending order, while discarding their positional indices. In this representation, only the activation strengths are preserved, and all positional information is removed.

We train logistic regression (LR) classifiers on sparse representation $\mathbf{z}$ to predict the binary accent label. 
Additionally, we train separate LR classifiers on $\mathbf{z}_\text{pos}$, and $\mathbf{z}_\text{mag}$ to evaluate the contribution of positional and magnitude information independently towards accent classification.
\vspace{-2mm}
\subsection{Interpretability measurement}
\label{subsec:explain_measure}
\vspace{-2mm}

We evaluate the accent classification F1 score  and use it as a task-level proxy for interpretability. However, the representation $\mathbf{u}$ derived from each NAC inherently encodes different amounts of accent-related information. Therefore, a higher F1 score does not imply higher interpretability. To enable fair comparison, in Table~\ref{table:referece}, we establish reference F1 scores for each NAC model $\mathcal{C}$ by training an LR classifier directly on $\mathbf{u}$. For each SAE configuration $(d_\text{z}, k)$, we compute the $ \Delta \text{F1}$ as:
\begin{align}
   \Delta \text{F1}^{\mathcal{C}}_{d_\text{z},k}
   = \text{F1}^{\mathcal{C}}_{d_\text{z},k}
   - \text{F1}^{\mathcal{C}}_{\text{ref}},
   \label{eq:delta}
\end{align}



A higher (i.e., less negative) $\Delta \text{F1}$ indicates that accent information remains largely preserved in the sparse representation $\mathbf{z}$, suggesting higher interpretability. 
Conversely, a larger performance drop implies that accent information is distributed across many dimensions and cannot be easily decomposed into sparse representations, indicating lower interpretability.

We use $\Delta \text{F1}$ computed on the full sparse activation $\mathbf{z}$ as the overall interpretability index for each NAC. 
In addition, $\Delta \text{F1}$ computed on the position-only features $\mathbf{z}_\text{pos}$ and magnitude-only features $\mathbf{z}_\text{mag}$ provide a more fine-grained analysis. It quantifies how much accent information is attributable to activation position and magnitude, respectively.


\vspace{-2mm}
\section{Experimental setup}
\label{sec:exp_setup}
\begin{figure*}[t]
    \centering
    \includegraphics[width=0.86\textwidth]{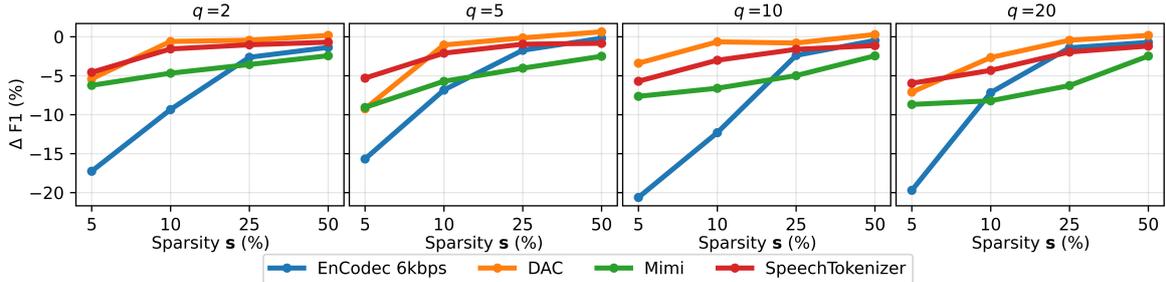}
    \vspace{-3mm}
    \caption{Interpretability measured by $\Delta$F1 (\%) under the US vs.\ UK setting. 
    Each subplot corresponds to a different latent ratio $q$. 
    The x-axis denotes sparsity level $s$, and higher $\Delta$F1 indicates better interpretability.}
    \label{fig:us_uk_result}
    \vspace{-5mm}
\end{figure*}

\begin{figure*}[t]
    \centering
    \includegraphics[width=0.86\textwidth]{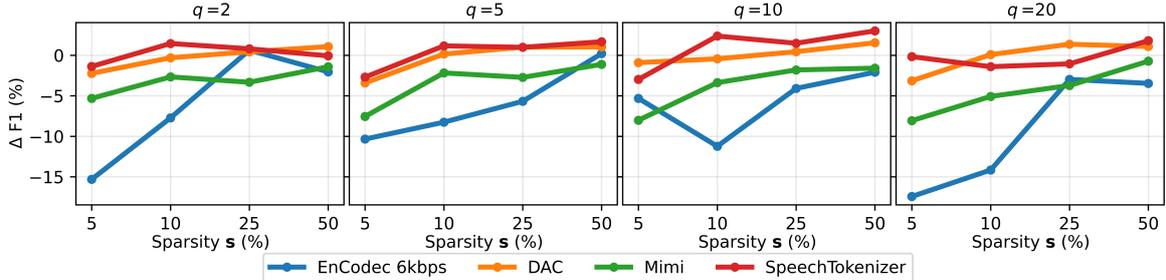}
    \vspace{-2mm}
    \caption{Interpretability measured by $\Delta$F1 (\%) under the US vs.\ Non-US-UK setting. 
    Each subplot corresponds to a different latent ratio $q$. 
The x-axis denotes sparsity level $s$, and higher $\Delta$F1 indicates better interpretability.}
    \label{fig:us_others_result}
    \vspace{-5mm}
\end{figure*}

\begin{table}[t]
    \centering
    \caption{Reference macro-F1 score (\%) of each NAC in US vs.\ UK and US vs.\ Non-US-UK settings. Higher value is better.}
    \vspace{-3mm}

\resizebox{\linewidth}{!}{
\begin{tabular}{l c c c c}
\toprule

& \shortstack{\textbf{EnCodec}\\ \textbf{1.5} / \textbf{6} / \textbf{12 kbps} } 
& \makecell{\textbf{DAC}}
& \makecell{\textbf{Mimi}}
& \shortstack{\textbf{Speech}\\\textbf{Tokenizer}} \\
\midrule

\textbf{US vs.\ UK}  
& 81.11 / 80.20 / 80.17
& 79.74
& 88.79 
& \textbf{92.44}\\

\midrule

\makecell[l]{\textbf{US vs.}\\\textbf{Non-US-UK}}
& 57.19 / 60.93 / 61.98
& 61.33
& 75.70
& \textbf{75.98} \\

\bottomrule
\end{tabular}
}
    \vspace{-8mm}
    \label{table:referece}
\end{table}

\vspace{-1mm}
\subsection{Dataset}
\label{subsec:dataset}
\vspace{-1mm}
\noindent \textbf{Vox-Profile.}
For both SAE and LR training (Sec.~\ref{subsec:SAE} and Sec.~\ref{subsec:LR}), we use the accent classification dataset described in the Vox-Profile benchmark~\cite{voxprofile}. Vox-Profile aggregates 11 open-source datasets with self-reported English accents and covers more than 16 English accents across different regions. 

\noindent \textbf{Task formulation and data split.}
We design two binary accent classification settings: \textbf{US vs.\ UK} and \textbf{US vs.\ Non-US-UK}.

In both settings, samples labeled as \textit{North America} are grouped as US accent (label = 0). 
For the US vs.\ UK setting, samples labeled as \textit{British Isles} in ~\cite{voxprofile}, excluding Irish, are grouped as UK accent (label = 1). 
For the US vs.\ Non-US-UK setting, samples labeled as regions outside \textit{North America} and the \textit{British Isles} (see Table 2 in~\cite{voxprofile}) are grouped as Non-US-UK (label = 1). 
We conduct experiments on both settings to ensure our findings generalize across accent distinctions.

We adopt the predefined speaker-disjoint train, validation, and test splits from Vox-Profile. 
In test sets, US accent samples account for 69.90\% and 63.98\% of samples in the US vs.\ UK and US vs.\ Non-US-UK settings, respectively.
To avoid data leakage between SAE and LR training, we divide the original validation set evenly into two non-overlapping subsets. 
The first half is used to monitor loss convergence in SAE (Sec.~\ref{subsec:SAE}), while the other half is used to train the LR classifier. 


\noindent \textbf{Audio preprocessing.}
For each audio sample, we crop the first 30 seconds, resample it to the sampling rate of the corresponding NAC, and convert it to the mono-channel format.

\vspace{-3mm}
\subsection{Neural audio codecs}
\label{subsec:nac_setup}
\vspace{-4mm}

\noindent \textbf{NAC selection.}
We include four widely used NAC models: EnCodec~\cite{encodec}, DAC~\cite{dac}, SpeechTokenizer~\cite{speechtokenizer}, and Mimi~\cite{moshi} in our experiments.
EnCodec and DAC are considered to yield representations that characterize acoustic properties. 
In contrast, SpeechTokenizer and Mimi capture stronger ``phonetic'' structure~\cite{ssl_phonetic}, as they distill representations from HuBERT~\cite{hubert} and WavLM~\cite{wavlm} during pre-training. While accent is partially reflected by acoustic properties like prosody and intonation, it is more closely related to phonetic characteristics~\cite{acc_syth}. Therefore, we hypothesize different accent information encoding behaviors between these two categories of NACs.



\noindent \textbf{NAC model configurations.}
EnCodec (24\,kHz, 1.5/6/12\,kbps, $d_{\text{codec}}=128$); 
DAC (24\,kHz, $d_{\text{codec}}=1024$); 
SpeechTokenizer (16\,kHz, $d_{\text{codec}}=1024$); 
Mimi (24\,kHz, $d_{\text{codec}}=512$).


\vspace{-2mm}
\subsection{Training \& hyperparameters}
\vspace{-1mm}
\label{subsec:sae_setup}
\noindent\textbf{SAE}: 
For SAE training, we adopt the official OpenAI TopK SAE implementation~\cite{topksae}. 
As the representation dimension $d_\text{codec}$ varies across different NACs, using a fixed latent dimension $d_\text{z}$ and sparsity level $k$ would lead to an unfair comparison. 
To ensure consistency across codecs, we scale $d_\text{z}$ proportionally to $d_\text{codec}$ with \textbf{latent ratio} $q=\frac{d_\text{z}}{d_\text{codec}} \in \{2, 5, 10, 20\}$. Value of $k$ is determined by a \textbf{relative sparsity} parameter $s \in \{0.5, 0.25, 0.10, 0.05\}$ where $k = \left\lceil s \cdot d_\text{codec} \right\rceil$, resulting in 16 total configurations. 
The relative sparsity parameter ensures each NAC representation retains the same proportion of active dimensions, avoiding bias from a fixed $k$ that could favor codecs with larger or smaller $d_{\text{codec}}$.


We train SAE models with batch size=1024 for 200 epochs. For configurations with $q \in \{2,5\}$ and $s \in \{0.10,0.05\}$, we extend training to 500 epochs due to slower convergence. Adam optimizer with a learning rate of $1e{-5}$ and $\epsilon = 6.25e{-10}$ is used. The source code is available at \href{https://github.com/Stanwang1210/Speech-SAE.git}{here}.

\noindent\textbf{Logistic regression}: 
For LR, we normalized input features and adopted \texttt{Saga} solver with L1 penalty and \texttt{max\_iter=3000}. Macro-F1 score is reported to account for class imbalance.

\vspace{-2mm}
\section{Result \& Analysis}
\label{sec:result}
\vspace{-1mm}
\subsection{NAC overall interpretability}
\label{subsec:result_overall}
\vspace{-2mm}
Figures~\ref{fig:us_uk_result} and~\ref{fig:us_others_result} show the $\Delta \text{F1}$ (Eq.~\ref{eq:delta}) under the US vs.\ UK and US vs.\ Non-US-UK settings, respectively. 

\noindent \textbf{US vs.\ UK.}
Across nearly all configurations, DAC consistently achieves the highest interpretability, ranking first in 13 out of 16 settings. SpeechTokenizer generally ranks second, while Mimi and EnCodec exhibit lower overall interpretability. 

A clear sparsity-dependent trend is observed. At very high sparsity ($s = 5\%$), all NACs yield negative $\Delta$F1 values, with EnCodec 6kbps showing the largest performance drop. As sparsity decreases ($s \ge 25\%$), $\Delta$F1 gradually approaches zero for all NACs. However, Mimi remains consistently below DAC and SpeechTokenizer, indicating that its accent information is less easily decomposed into sparse dimensions.

\noindent \textbf{US vs.\ Non-US-UK.}
Here, SpeechTokenizer shows the highest interpretability, ranking first $\Delta \text{F1}$ in 14 out of 16 configurations. The sparsity trend remains similar: at high sparsity, EnCodec exhibits the largest degradation, whereas SpeechTokenizer and DAC perform relatively stable. As sparsity decreases, $\Delta \text{F1}$ approaches zero or becomes positive for SpeechTokenizer, indicating accent information can be decomposed into sparse dimensions and entangled information is removed.

Overall, although phonetic-oriented NACs (Mimi and SpeechTokenizer) retain more raw accent information than acoustic-oriented NACs (EnCodec and DAC), as shown in Table~\ref{table:referece}, it does not necessarily imply higher interpretability. 

\vspace{-2mm}
\subsection{Analysis: position vs.\ magnitude}
\label{subsec:analysis_position_value}
\vspace{-1mm}
Table~\ref{tab:delta_position_value} reports the $\Delta$F1 (\%) of training LR on the full SAE activation $\mathbf{z}$, and its derived position-only ($\mathbf{z}_\text{pos}$) and magnitude-only ($\mathbf{z}_\text{mag}$) features for $s=5\%$ in US vs.\ UK setting. 
We focus on $s=5\%$ as it exhibits the largest discrepancy across NACs.
From Table~\ref{tab:delta_position_value}, we draw the following observations:

\noindent\textbf{Acoustic NACs encode accent information primarily in activation magnitude.}  
For acoustic-oriented NACs, the $\Delta$F1 from magnitude-only features is generally better than that from position-only features. 
This indicates for acoustic NACs, accent information is largely reflected in the strength of sparse activations rather than their positions of activations.

\noindent\textbf{Phonetic NACs encode accent information primarily in activation position.}  
In contrast, for phonetic-oriented NACs, magnitude-only features result in more negative $\Delta$F1 compared to position-only features.
This suggests that accent information in phonetic NACs is mainly encoded in the position of activated dimensions rather than their magnitudes.

\begin{figure}[t]
    \centering
    \includegraphics[width=0.9\linewidth]{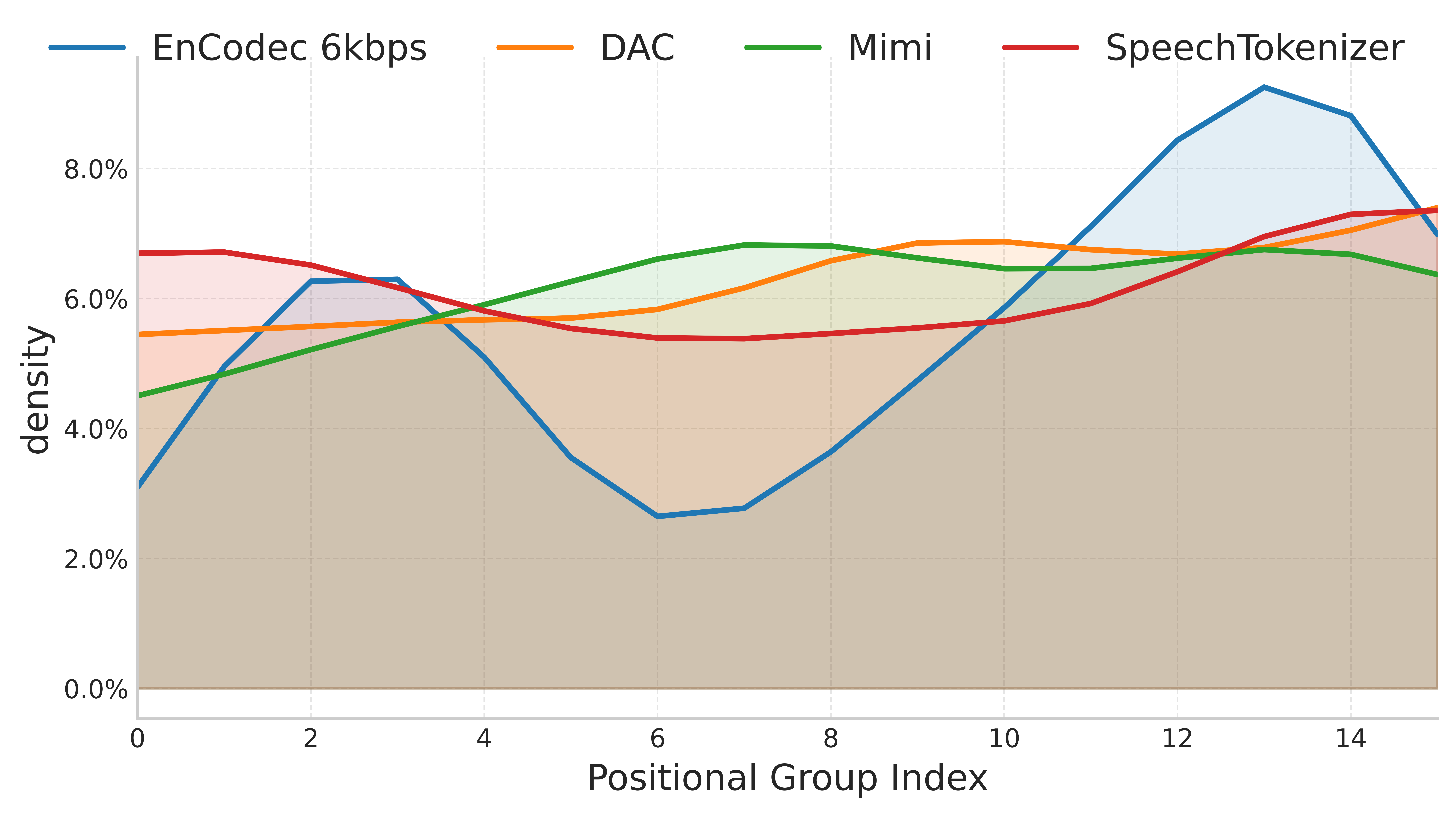}
    \vspace{-4mm}
    \caption{Activation density distribution across ordered latent dimensions on the test set. Dimensions in $\mathbf{z}$ are divided into 16 sequential groups; the x-axis shows group index and the y-axis shows activation density. }
    \label{fig:inds_distribution}
    \vspace{-7mm}
\end{figure}

Among all NACs, EnCodec-6kbps exhibits the most negative $\Delta$F1 when using position-only features, suggesting that the position may not be informative for accents. We hypothesize that for EnCodec-6kbps, the SAE struggles to decompose information into distinct dimensions, making the position indices less discriminative. To investigate this, in Fig.~\ref{fig:inds_distribution}, we present the activation density distribution over the latent space of $\mathbf{z}$ on the test set when $q$=2 and $s$=0.05, where $d_\text{z}$ is grouped into 16 ordered groups for clarity. We observe that EnCodec-6kbps shows a notably non-uniform activation distribution compared to other NACs, suggesting poor decomposition across dimensions. Consequently, results from position-only features suffer degradation due to interference from entangled information.


\begin{table}[t]
\centering
\caption{Comparison of information encoding in the US vs.\ UK setting across NACs and $q$ when $s=5\%$. 
$\Delta$F1 (\%) is reported for full SAE activations $\mathbf{z}$ (first row of each section), position-only features $\mathbf{z}_\text{pos}$ (second row), and magnitude-only features $\mathbf{z}_\text{mag}$ (third row). 
Higher value indicates better interpretability.}
\vspace{-3mm}
\resizebox{0.9\linewidth}{!}{
\begin{tabular}{lcccc}
\toprule
\multicolumn{1}{c}{\textbf{Latent Ratio $q$}} & 2 & 5 & 10 & 20 \\
\midrule
\multicolumn{1}{l}{\textbf{EnCodec 6kbps}} & -17.25 & -15.64 & -20.37 & -19.70 \\
\hspace{1em}Position  & -39.06 & \textbf{-19.16} & -38.90 & -45.06 \\
\hspace{1em}Magnitude & \textbf{-19.64} & -25.82 & \textbf{-25.58} & \textbf{-20.67} \\
\midrule
\multicolumn{1}{l}{\textbf{DAC}} & -5.59 & -9.46 & -3.36 & -7.29 \\
\hspace{1em}Position  & -9.80 & -12.04 & -9.88 & -10.93 \\
\hspace{1em}Magnitude & \textbf{-5.00} & \textbf{-5.08} & \textbf{-4.78} & \textbf{-3.32} \\
\midrule
\multicolumn{1}{l}{\textbf{Mimi}} & -6.43 & -9.72 & -7.69 & -8.82 \\
\hspace{1em}Position  & \textbf{-8.12} & \textbf{-10.49} & \textbf{-8.64} & \textbf{-9.27} \\
\hspace{1em}Magnitude & -38.87 & -19.02 & -24.70 & -22.46 \\
\midrule
\multicolumn{1}{l}{\textbf{SpeechTokenizer}} & -4.54 & -5.21 & -5.83 & -6.33 \\
\hspace{1em}Position  & \textbf{-7.20} & \textbf{-7.09} & \textbf{-6.76} & \textbf{-7.20} \\
\hspace{1em}Magnitude & -23.82 & -29.38 & -23.62 & -27.03 \\
\bottomrule
\end{tabular}
}
\vspace{-7mm}
\label{tab:delta_position_value}
\end{table}

\begin{table}[t]
\centering
\caption{Bitrate analysis of EnCodec models in the US vs.\ UK setting. 
$\Delta$F1 (\%) is reported under different latent ratio $q$ and sparsity $s$. 
Higher value indicates stronger interpretability.}
\vspace{-2mm}
\resizebox{\linewidth}{!}{
\begin{tabular}{ccccccc}
\toprule
& \multicolumn{2}{c}{1.5 kbps} 
& \multicolumn{2}{c}{6 kbps} 
& \multicolumn{2}{c}{12 kbps} \\
$q$ 
& $s=5\%$ & $10\%$ 
& $s=5\%$ & $10\%$  
& $s=5\%$ & $10\%$  \\
\cmidrule(lr){1-1}\cmidrule(lr){2-3}\cmidrule(lr){4-5}\cmidrule(lr){6-7}
2  &\textbf{ -7.25} &\textbf{ -2.52} & -17.25 & -9.28 & -17.78 & -13.69 \\
5  &\textbf{ -7.05} & \textbf{-2.91} & -15.64 & -6.81 & -16.51 & -11.86 \\
10 & \textbf{-9.66} & \textbf{-5.00} & -20.37 & -12.29 & -16.30 & -9.06 \\
20 & \textbf{-8.72} & \textbf{-4.40} & -19.70 & -6.94 & -23.18 & -8.06 \\
\bottomrule
\end{tabular}
}
\label{tab:bitrate_comparison}
\vspace{-7mm}
\end{table}

\vspace{-1mm}
\subsection{Analysis: bitrate}
\vspace{-1mm}
\label{subsec:analysis_bitrate}

To explore the impact of bitrate on interpretability of accent information, we perform comparison between EnCodec under different bitrates (1.5, 6 and 12 kbps) for US vs.\ UK setting.
From Table~\ref{tab:bitrate_comparison}, while Encodec models across different bitrates provide similar reference performance (Table~\ref{table:referece}), their interpretability levels decrease noticeably as bitrate increases.
In particular, the 1.5\,kbps model exhibits the smallest performance drop (i.e., $\Delta$F1 $< 10\%$), indicating higher interpretability. One possible explanation is that the lower-bitrate model relies on fewer codebooks, which may make accent information easier to decompose. We will explore how bitrate affects interpretability in other NACs in future work.
\vspace{-2mm}
\section{Conclusion, Limitations, \& Future work}
\vspace{-1mm}

In this work, we propose a framework to quantify the task-level interpretability of Neural Audio Codecs (NACs) and demonstrate its effectiveness using a representative paralinguistic feature--accent as a case study. We employ Sparse Autoencoders (SAEs) to extract interpretable sparse activations from NAC representations and further decompose them into position and magnitude representations for detailed analysis. We evaluate four NAC models under 16 SAE configurations, using $\Delta$F1 as an interpretability index. Our results show that DAC and SpeechTokenizer achieve the highest interpretability. We observe that acoustic-oriented NACs primarily encode accent information in activation magnitudes, whereas phonetic-oriented NACs rely more on activation positions. Finally, we find that lower-bitrate variant of EnCodec exhibits higher interpretability than its higher-bitrate counterparts.

This work provides a foundation for future research in NAC interpretability. We note that our framework is intended to evaluate task-level interpretability. We also acknowledge that absolute sparse capacity may affect comparisons involving lower-dimensional NACs such as EnCodec. In the future, we plan to extend this framework to more NAC models, investigate phenomena such as feature splitting and feature steering, and explore interpretability across other paralinguistic attributes.

\clearpage
\section{Acknowledgments}
This work was supported by the Office of the Director of National Intelligence (ODNI), Intelligence Advanced Research Projects Activity (IARPA), via the ARTS Program under contract D2023-2308110001. The views and conclusions contained herein are those of the authors and should not be interpreted as necessarily representing the official policies, either expressed or implied, of ODNI, IARPA, or the U.S. Government. The U.S. Government is authorized to reproduce and distribute reprints for governmental purposes notwithstanding any copyright annotation therein.




\section{Generative AI Use Disclosure}

We used ChatGPT-5.2 to assist with polishing the writing of this paper. All substantive content and contributions of the manuscript were developed by the authors.




\bibliographystyle{IEEEtran}
\bibliography{mybib}

\end{document}